\documentclass[10pt,twocolumn]{article}
\usepackage{graphicx}
\usepackage{ol}
\usepackage{amsmath}
\usepackage{times}
\usepackage{mathptmx}
\usepackage{psfrag}

\newcommand{\parti}[2]{\frac{\partial #1}{\partial #2}}
\newcommand{\partit}[2]{\frac{\partial^2 #1}{\partial #2^2}}

\newcommand{\etal}{\textit{et al.}\ }
\begin{document}

\twocolumn[

\title{Metaphoric optical computing of fluid dynamics}
\author{Mankei Tsang and Demetri Psaltis}
\date{\today}
\affiliation{
Department of Electrical Engineering, 
California Institute of Technology, Pasadena, California 91125}
\begin{abstract}
We present theoretical and numerical evidence to show that
self-defocusing nonlinear optical propagation can be used to compute
Euler fluid dynamics and possibly Navier-Stokes fluid dynamics. In
particular, the formation of twin vortices and the K\'arm\'an vortex
street behind an obstacle, two well-known viscous fluid phenomena, is
numerically demonstrated using the nonlinear Schr\"odinger equation.
\end{abstract}
]

\section{Introduction}
\subsection{Philosophy of metaphoric computing}
Nonlinear dynamical systems, such as weather, plasma, and the economy,
are ubiquitous in nature and everyday's life, yet such
systems are typified by their highly complex and chaotic behaviors,
making them notoriously difficult to study theoretically,
experimentally, and numerically.  Analytic solutions of nonlinear
systems are rare, experiments are often too inflexible or impractical,
and numerical simulations must take into account a large number of
data points in multiple dimensions in order to accurately model a
problem of interest, such that even the fastest supercomputers today
would take days or weeks to simulate relatively simple nonlinear
dynamics that a physical system exhibits in seconds.

On the opposite side of the same coin, we can regard the physical
system as a computational device that computes its own dynamics at a
speed unimaginable by supercomputers. The key to harnessing this
tremendous computing power of a physical system is therefore to make
it compute other interesting problems of the same order of
complexity. Of course, a conventional digital computer is itself a
physical system, but it makes use of complex semiconductor physics to
compute elementary logic operations, and in doing so, discards a large
amount of information that is considered extraneous.  In this
perspective, a digital computer is an extremely inefficient computing
device, as it only utilizes an exceedingly small amount of the full
computing capability potentially offered by its physics. The advantage
in this case is the flexibility in cascading different logic
operations for general-purpose computing, but as evidenced by the
difficulties in the numerical simulations of nonlinear dynamical
systems, this inefficient computing method is often inadequate.

In order to make full use of the computing capability offered by a
physical system, we hereby propose the concept of metaphoric
computing, which makes use of a more experimentally accessible
nonlinear dynamical system to simulate another nonlinear dynamical
system. An example of this computing method is a wind tunnel, in which
a small-scale fluid experiment is performed to simulate large-scale
fluid dynamics, by virtue of the scaling laws inherent in fluid
dynamics.  Metaphoric computing, however, is not restricted to the use
of similar physical systems to simulate each other. In this paper, we
show in particular that nonlinear optics can compute fluid dynamics as
well.  An optical beam inherently holds three-dimensional
spatiotemporal information, and nonlinear
optical propagation computes the evolution of this large amount of
information simultaneously at the speed of light, promising
substantial parallelism and speed for computing. Although the use of
nonlinear optics for digital computing has not been as successful
as the use of solid-state electronics, forcing optical beams to
compute binary logic wastes most of the spatiotemporal information
that can be manipulated in optical beams. Instead of fitting a square
peg in a round hole, using optics to simulate other nonlinear
dynamical systems provides a natural way of making full use of the
computing capacity offered by a nonlinear optical system.

Fluid dynamics, the foundation of a wide variety of important research
fields including meteorology, aeronautics, plasma physics,
superfluids, and Bose-Einstein condensates, is an ideal problem to
solve by metaphoric computing. Intractable theoretical analysis and
inflexible experiments compel the use of numerical simulations, the
difficulty of which nonetheless gives rise to a whole new field,
computational fluid dynamics, in itself.  The main difficulty is due
to the inherent complexity of a fluid dynamics problem, which is
nonlinear and continuously generates finer structures as the fluid
dynamics evolves. For problems that are of practical interest, such
fine structures are often orders-of-magnitude smaller than the size of
the objects under consideration, thus requiring a large number of data
points in each of the three spatial dimensions to be manipulated at each time
step, which must also be correspondingly small to avoid numerical
instabilities.  An alternative method of simulating complex fluid
dynamics that combines the speed of a fluid experiment and the
flexibility of a numerical analysis is hence of great practical
importance. In this paper, we show that, via a suitable
transformation, nonlinear optical propagation can be utilized to
simulate Euler fluid dynamics, which is known to be computationally
expensive and unstable to solve numerically . We also provide strong
evidence that nonlinear optics can simulate high-Reynolds-number
Navier-Stokes fluid dynamics as well, which include a large class of
important and computationally difficult problems, such as
turbulence. With the speed, parallelism, and configurability of
optics, an ``optical wind tunnel'' may one day become a viable
alternative to experiments and numerical analysis in the study of
fluid dynamics.

\subsection{Correspondence between nonlinear optics
and fluid dynamics}
The analogy between nonlinear optics and fluid dynamics has been noted
by many authors
\cite{wagner,coullet,brambilla,arecchi,akhmanov,swartzlander,staliunas,vaupel,molina,roux,rozas_PRA,rozas_PRL,michinel,paz-alonso,paz-alonso_PRL,pomeau,bolda,chiao}.
Wagner \etal first suggested that the nonlinear propagation equation
of an optical beam can be recast into equations that resemble the
continuity equation and the Bernoulli equation in irrotational fluid
dynamics \cite{wagner}.  Coullet \etal first coined the term ``optical
vortices,'' which shows the analogy between phase singularities in
optics and fluid vortices \cite{coullet}. Brambilla \etal noted that
laser equations can be transformed to a hydrodynamic form
\cite{brambilla}.  Arecchi \etal first experimentally demonstrated the
dynamics of optical vortices in nonlinear optics \cite{arecchi}.
Akhmanov \etal called the rich nonlinear dynamics observed in a
nonlinear resonator ``optical turbulence.'' \cite{akhmanov}
Swartzlander and Law observed optical vortex solitons created via the
instability of dark soliton stripes analogous to the Kelvin-Helmholtz
instability in fluid dynamics \cite{swartzlander}. Staliunas showed
that a laser can be described by the Ginzburg-Landau equation, which
can be transformed into equations resembling the Navier-Stokes
equations that describe viscous fluid dynamics \cite{staliunas}.
Vaupel \etal observed vortex pair nucleation by the interference of
two modes in a laser and claimed that it was an analog of a vortex
street behind an obstacle in a fluid flow
\cite{vaupel}. Molina-Terriza \etal also observed optical vortex
streets in walking second-harmonic generation \cite{molina}.  Roux
\cite{roux} and Rozas \etal \cite{rozas_PRA} studied the interactions
between optical vortices and found that their interactions resemble
those of fluid vortices.  Rozas \etal then experimentally demonstrated
the fluidlike motion of a pair of optical vortices \cite{rozas_PRL}.
Michinel \etal \cite{michinel} and Paz-Alonso \etal \cite{paz-alonso}
found that optical propagation in a cubic-quintic nonlinear medium
resembles a liquid drop, and optical vortices in such a medium also
have fluidlike motions \cite{paz-alonso_PRL}.  On the other hand,
nonlinear optics has been compared with superfluids and Bose-Einistein
condensates, as they can all be described, to varying degrees, by the
nonlinear Schr\"odinger equation \cite{boyd,agrawal}, commonly known
as the Gross-Pitaevskii equation in the field of superfluids
\cite{donnelly}.  Pomeau and Rica suggested that the phenomenon of
transition to dissipation in a superflow \cite{frisch} can be observed
in nonlinear diffraction \cite{pomeau}. Bolda \etal numerically
demonstrated the same phenomenon in a nonlinear Fabry-P\'erot cavity
\cite{bolda}. Chiao also found that photons in such a cavity should
obey the Bogoliubov dispersion relation for a superfluid \cite{chiao}.

The abundant amount of prior work credited above provides ample evidence
that nonlinear optics resembles fluid dynamics to a certain degree. In
order to use nonlinear optics as a useful and practical computational
tool for fluid dynamics, however, simply drawing analogies between the
two kinds of dynamics is not enough. One must be able to show an exact
correspondence, or at the very least, an approaching convergence
between a problem in nonlinear optics and a problem in fluid dynamics,
in order to produce any useful prediction of fluid dynamics via
nonlinear optics. Moreover, as computers nowadays have enough
capabilities to simulate two-dimensional fluids, the mere
correspondence between optics and two-dimensional fluid dynamics
considered in most of the prior work would not motivate the use of
metaphoric optical computing in preference to conventional digital
computing. A three-dimensional fluid modeling, on the other hand,
requires a processing capability orders-of-magnitude higher than that
available in today's supercomputers, so metaphoric optical computing
would need to compute such problems much more efficiently to
compete with electronic computers and the Moore's law.

In the following sections, we shall attempt to establish the
correspondence between nonlinear optics and three-dimensional fluid
dynamics. We shall show that, taking group-velocity dispersion into
account, nonlinear optical dynamics approaches three-dimensional
inviscid Euler fluid dynamics in the highly nonlinear self-defocusing
regime, where the optical intensity represents the fluid density, the
optical phase gradient represents the fluid velocity, the nonlinear
refractive index perturbation represents pressure, the propagation
distance represents time, and the temporal dimension of the optical
pulse represents the third dimension of the fluid. As Euler fluid
equations often exhibit high numerical instabilities, this
correspondence in itself should be useful in modeling
high-Reynolds-number fluid dynamics away from objects and boundaries.
In the convergence of nonlinear Schr\"odinger equation towards the
Euler equations, a ``quantum pressure'' term arising from the
nonlinear Schr\"odinger equation plays the role of a small
parameter. As this quantum pressure term plays analogous roles to
viscosity in the Navier-Stokes equations, we argue that nonlinear
optics should be able to approximate viscous Navier-Stokes fluid
dynamics as well, in the regime where quantum pressure and viscosity
both play the role of small parameters in the respective
equations. That said, we do not pretend that we have established the
equivalence between nonlinear optics and Navier-Stokes dynamics, as
the similarity between quantum pressure and viscosity is still an open
problem.

On the practical side, in cases where ideal nonlinear optics setup
is not available, we suggest a split-step method that pieces together
different optical devices to approximate an ideal nonlinear optics
experiment. This method is very similar to the method proposed to
simulate quantum systems using a quantum computer \cite{lloyd}.

It must be stressed that although we focus on simulations of classical
physical systems, future quantum computers that simulate quantum
systems \cite{lloyd} would run into the same problem of manipulating a
large amount of multi-dimensional information. In the case of quantum
systems, multi-dimensional quantum information, such as a
multi-particle multi-spatiotemporal-dimensional wavefunction, needs to
be processed in parallel. Quantum computers can naturally parallelize
the multi-particle aspect, but there is no obvious way of
parallelizing the manipulation of multi-spatiotemporal-dimensional
information via simple binary quantum logic. Perhaps a quantum
metaphoric computing would then be necessary, where a more accessible
multi-dimensional quantum system is used to simulate another quantum
system.

\section{Correspondence between nonlinear optics
and Euler fluid dynamics}
\subsection{Madelung transformation}
We now proceed to show mathematically how the self-defocusing
nonlinear optical propagation equation, including the effect of
group-velocity dispersion, can be transformed to three-dimensional
hydrodynamic equations. First, we show how the optics equations,
in the absence of optical vortices, correspond to inviscid and
irrotational fluid equations. This form of transformation is widely
attributed to Madelung \cite{madelung}.  We model the paraxial
nonlinear propagation of an optical beam, described by the envelope
function $\psi(z,x,y,T)$, via the nonlinear Schr\"odinger equation
\cite{boyd,agrawal},
\begin{align}
i\parti{\psi}{z} &=
-\frac{1}{2k_0}\left(\partit{}{x}+\partit{}{y}\right)\psi
+\frac{\beta_2}{2}\partit{\psi}{T}
-k_0 n_2|\psi|^2 \psi,
\label{nls}
\end{align}
where $z$ is the propagation distance, $k_0 = 2\pi n_0/\lambda_0$ is
the carrier wave number, $\beta_2$ is the group-velocity dispersion
coefficient, $T$ is the time coordinate in the moving frame of the
pulse, and $n_2$ is the nonlinear Kerr coefficient. To use the
time coordinate as the third spatial dimension of a fluid, anomalous
group-velocity dispersion, or $\beta_2< 0$, is required.
 Dispersion can then be regarded in equal footing as
diffraction if a normalized time coordinate is defined as
\begin{align}
\tau \equiv \frac{T}{\sqrt{-\beta_2n_0k_0}},
\end{align}
such that
\begin{align}
i\parti{\psi}{z} &=
-\frac{1}{2k_0}\left(\partit{}{x}+\partit{}{y}+\partit{}{\tau}\right)\psi
-k_0 n_2|\psi|^2 \psi.
\end{align}
The Madelung transformation is defined as follows,
\begin{align}
\psi &= |\psi|\exp(j\phi),
\\
I &= |\psi|^2,
\\
\mathbf{k} &= \nabla'\phi =
\hat{x}\parti{}{x}+\hat{y}\parti{}{y}+\hat{\tau}\parti{}{\tau},
\end{align}
such that the evolution equations for the intensity, $I$, and the
phase gradient, $\mathbf{k}$, are given by
\begin{align}
\parti{I}{z} + \frac{1}{k_0}\nabla'\cdot(I\mathbf{k}) &= 0,
\label{continuity}\\
\parti{\mathbf{k}}{z} +
\frac{1}{k_0}\nabla'\left(\frac{1}{2}\mathbf{k}\cdot\mathbf{k}\right)
&= \nabla'(k_0 n_2 I) + 
\nonumber\\&\quad
\frac{1}{k_0}\nabla'\left(\frac{1}{2\sqrt{I}}
\nabla'^2\sqrt{I}\right).
\label{bernoulli}
\end{align}
One can already see that Eq.~(\ref{continuity}) has the exact same form
as the fluid continuity equation, while Eq.~(\ref{bernoulli}) resembles the
Bernoulli equation \cite{wagner}, if one regards the intensity as
the fluid density and the phase gradient as the fluid velocity.
 The nonlinear refractive index term,
$k_0n_2I$, would resemble the fluid pressure if $n_2<0$, so self-defocusing
is required. The last term in Eq.~(\ref{bernoulli}) is a peculiar term that
arises from optical diffraction and dispersion,
does not exist in classical fluid dynamics, and is commonly
called the ``quantum pressure.''

In order to compare these equations
with fluid equations more easily, we use the following normalized
variables,
\begin{align}
\nabla = W\nabla', \quad \zeta = \frac{K}{Wk_0}z,
\\
\rho = \frac{I}{I_0},
\quad
\mathbf{u} = \frac{\mathbf{k}}{K}=\frac{1}{KW}\nabla\phi,
\label{defn}
\\
a = \frac{1}{k_0\sqrt{-n_2 I_0}}, \quad \mathcal{M} = Ka,
\quad \mathcal{R} = KW,
\end{align}
where $W$ is the characteristic size, $K$ is the characteristic phase
gradient, $I_0$ is some characteristic optical intensity of the
propagation, and $a$ is the so-called ``healing'' length, which is the
length scale at which the quantum pressure term has the same order of
magnitude as the nonlinear term on the right hand side of
Eq.~(\ref{bernoulli}), $\mathcal{M}$ is the Mach number, which
measures the relative strength of fluid pressure compared with
convection, and $\mathcal{R}$ is another number that measures the
relative strength of fluid convection compared with quantum
pressure. The normalized equations become
\begin{align}
\parti{\rho}{\zeta} + \nabla\cdot(\rho\mathbf{u}) &= 0,
\label{norm_continuity}\\
\parti{\mathbf{u}}{\zeta}+
\nabla\left(\frac{1}{2}\mathbf{u}\cdot\mathbf{u}\right)
&= -\frac{1}{\mathcal{M}^2\rho}\nabla\left(\frac{1}{2}\rho^2\right)
\nonumber\\&\quad
-\frac{1}{\mathcal{R}^2}
\nabla\left(\frac{1}{2\sqrt{\rho}}\nabla^2\sqrt{\rho}\right).
\label{norm_bernoulli}
\end{align}
Equation (\ref{norm_continuity}) is exactly the same as the fluid
continuity equation, and in the limit of $\mathcal{M}/\mathcal{R}\to
0$, which is the highly self-defocusing regime,
Eq.~(\ref{norm_bernoulli}) is the same as the hydrodynamic equation of
motion that describes inviscid and irrotational fluids. Equations
(\ref{norm_continuity}) and (\ref{norm_bernoulli}) also admit sound
wave solutions, which describe travelling perturbations to the density
and the velocity.  As long as the sound waves are weak, the dependence
of pressure on the density is not crucial, and the use of
self-defocusing Kerr nonlinearity is adequate. This restricts the
correspondence to slightly compressible barotropic fluids.

In order to model slightly compressible fluids, the optical beam needs
to have a relatively constant intensity background. This can be
achieved approximately near the center of a very large beam, in a
large multimode waveguide as a container in two spatial dimensions, or in a
cubic-quintic nonlinear medium to provide a ``surface tension'' to the
beam \cite{quiroga,michinel,paz-alonso,paz-alonso_PRL}.

\subsection{Vorticity}
In general, the fluid velocity vector should contain an irrotational
component and a rotational component,
\begin{align}
\mathbf{u} &= -\nabla\varphi -\nabla\times\mathbf{A},
\end{align}
where $\varphi$ is called the velocity potential,
and the curl of $\mathbf{u}$ is defined as the fluid vorticity,
\begin{align}
\pmb\omega &= \nabla\times\mathbf{u} = -\nabla\times(\nabla\times\mathbf{A}).
\end{align}
The dynamics of vorticity is arguably the cornerstone of hydrodynamics
\cite{donnelly}.  The inviscid fluid dynamics that includes the
rotational effect is governed by the Euler equation,
\begin{align}
\parti{\mathbf{u}}{\zeta}+
\mathbf{u}\cdot\nabla\mathbf{u}
&= -\frac{1}{\mathcal{M}^2\rho}\nabla P,
\label{euler_bernoulli}
\end{align}
where $P$ is the pressure. For incompressible fluids,
$\mathcal{M} << 1$, and as long as $P$ increases with $\rho$, the
specific dependence of $P$ on the fluid properties is not
important. Equation (\ref{euler_bernoulli}) contains the convective
term $\mathbf{u}\cdot\nabla\mathbf{u}$, which can be written as
\begin{align}
\mathbf{u}\cdot\nabla\mathbf{u} &= 
\nabla\left(\frac{1}{2}\mathbf{u}\cdot\mathbf{u}\right)+
(\nabla\times\mathbf{u})\times\mathbf{u}
\\
&=\nabla\left(\frac{1}{2}\mathbf{u}\cdot\mathbf{u}\right)+
\pmb\omega\times\mathbf{u}.
\end{align}
One can then see that the optical Bernoulli equation,
Eq.~(\ref{norm_bernoulli}), misses the rotational component
of the convective term. In other words, the Madelung
transformation is only able to describe the irrotational
part of the fluid motion, but not the more important rotational
part.

The inability of the Madelung transformation to describe vorticity
is due to the failure of the transformation near optical vortices, where
Eq.~(\ref{norm_bernoulli}) is ill-defined. To understand
this problem, consider a rectilinear optical vortex
in polar coordinates and neglect the third fluid dimension for now,
\begin{align}
\psi = f(r)\exp(im\theta),\label{polar}
\\
r = \sqrt{x^2+y^2}, \quad \theta = \tan^{-1}\left(\frac{y}{x}\right),
\end{align}
where $m$ is an integer and is called the topological charge of an optical
vortex. The phase gradient is then given by
\begin{align}
\mathbf{k} &= \hat{\theta}\frac{1}{r}\parti{}{\theta}(m\theta) =
\hat{\theta}\frac{m}{r}.
\end{align}
The fluid vorticity is proportional to the curl of $\mathbf{k}$,
\begin{align}
\nabla\times\mathbf{k} &= \hat{z}\frac{2\pi m}{r}\delta(r),
\label{pointvortex}
\end{align}
which resembles the vorticity of an ideal point fluid vortex
\cite{thomson}. The motion of these vortices, however, cannot be
described by the Madelung transformed equations due to two problems:
$\mathbf{k}$ diverges when $r \to 0$, so the fluid velocity
$\mathbf{u}$ at the center of a vortex is infinite, 
and $f(r)$ must approach $r^m$ in the limit of $r\to 0$
to maintain the continuity of $\psi$, so the quantum pressure term,
with $\sqrt{\rho}$ in the denominator, is also infinite near the
vortex center.

To overcome these difficulties, it is necessary to consider the motion
of the optical vortices separate from the irrotational optical flow.

\subsection{Optical vortex solitons and point vortices}
In a relatively constant intensity background, optical vortices exist
as optical vortex solitons \cite{snyder,swartzlander,quiroga}. The
optical envelope function $\psi$ of a vortex soliton is given by
Eq.~(\ref{polar}), where $f(r)\to r^m$ for $r << a$, and $f(r)$
approaches a constant for $r >> a$, where $a$ is the healing length
and also the size of the dark spot of a vortex soliton. In three
dimensions, a vortex soliton exists as a vortex filament. We shall
hereafter consider single-charged vortex solitons with $m = \pm 1$
only, as they have the lowest energy and are the most prevalent
ones arising from an experimental situation. It is also
more accurate to approximate continuous vorticity with only
discrete vortices with the smallest topological charge.

\begin{figure}[htbp]
\centerline{\includegraphics[width=0.45\textwidth]{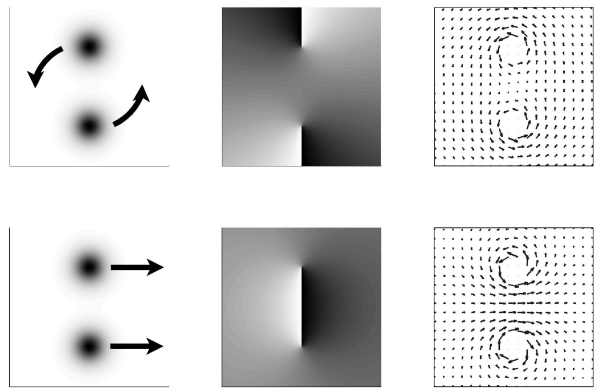}}
\caption{Intensity (left column), phase (middle
column), and phase gradient (right column)
of two optical vortex solitons with the same charge (top row),
which should rotate in the same sense, and those of two
vortex solitons with opposite charges, which should drift in
a direction perpendicular to their separation. The phase gradient
near the centers of the vortices is not plotted due to its divergence.}
\label{two_vortices}
\end{figure}

Eq.~(\ref{pointvortex}) suggests that an optical vortex soliton
resembles an ideal point vortex in incompressible fluids. Indeed, the
motion of optical vortices in the highly self-defocusing limit can be
rigorously proven to behave in the same way as point fluid vortices
\cite{roux,rozas_PRA,neu,pismen,lin,lin2}. If one defines the position of
each vortex filament as $\mathbf{x}_j$, then the fluid velocity at
each point due to the presence of the vortex filaments in the
limit of high self-defocusing is given by
\begin{align}
\mathbf{u}(\mathbf{x},\zeta)
&= -\sum_j 2\pi m_j \int
\frac{(\mathbf{x}-\mathbf{x}_j)
\times d\mathbf{x}_j}
{4\pi|\mathbf{x}-\mathbf{x}_j|^3}-\nabla\varphi,
\quad a \to 0,
\end{align}
where $\mathbf{x}$ is the normalized three-dimensional position
vector, $m_j$ is the topological charge of vortex $j$, and
$-\nabla\varphi$ describes the irrotational flow according to
Eq.~(\ref{norm_bernoulli}). In particular, the motion of each filament
is given by,
\begin{align}
\parti{\mathbf{x}_i}{\zeta} &= -\sum_{j} 2\pi m_j \int
\frac{(\mathbf{x}_i-\mathbf{x}_j)
\times d\mathbf{x}_j}
{4\pi|\mathbf{x}_i-\mathbf{x}_j|^3}-\nabla\varphi,
\quad
a \to 0,
\end{align}
in the leading order. These equations of vortex motion are valid as
long as the separations of the vortices are much larger than $a$. For
example, Fig.~\ref{two_vortices} plots the intensity, phase, and
phase gradient of two rectilinear optical vortex solitons with the
same charge, which should rotate around each other, and
those of two vortices with opposite charges, which should drift in
the same direction perpendicular to their separation.

With the vortex filaments, one can define the equivalent vorticity in
an optical beam,
\begin{align}
\pmb\omega(\mathbf{x},\zeta) &= \sum_j 2\pi m_j \int d\mathbf{x}_j
\delta(\mathbf{x}-\mathbf{x}_j),
\end{align}
which can be used to approximate the continuous vorticity of a fluid,
if the number of vortex filaments is large enough. In this case,
to include the vorticity effect in the nonlinear optical dynamics,
one can phenomenologically patch up the irrotational equation of motion,
Eq.~(\ref{norm_bernoulli}), 
\begin{align}
\parti{\mathbf{u}}{\zeta} +
\nabla\left(\frac{1}{2}\mathbf{u}\cdot\mathbf{u}\right)
+\pmb\omega\times\mathbf{u} &= -\frac{1}{\mathcal{M}^2}\nabla\rho.
\end{align}
This modification of the equation of motion can be attributed to the
phenomenon of phase slippage \cite{donnelly,sonin}, well known in the
field of superfluids.  The use of discrete point vortex interactions
to calculate Euler fluid dynamics is also a well-known numerical
method in computational fluid dynamics \cite{leonard}. Hence, to
simulate Euler fluid dynamics, one can approximate both the rotational
and irrotational components of the initial fluid velocity profile by
the optical phase and the phase singularities in an optical beam, and
the nonlinear self-defocusing propagation of the beam would converge
to incompressible Euler fluid dynamics in the strongly self-defocusing
regime. One can also borrow from the well-established numerical
techniques \cite{leonard} to determine how the distribution of
optical vortices sufficiently approximates the continuous vorticity
in fluids.

\subsection{The fluid flux representation}
So far, we have shown that optical vortex solitons behave like point
vortices in fluids when they are far away from each other, and this
behavior can be used to approximate Euler fluid dynamics.  However,
there is no guarantee that the vortices would remain well separated in
the course of the vortex dynamics. If optical vortices behaved exactly
like point vortices, then their velocities would diverge when they are
close to each other. This velocity divergence is well known to cause
significant numerical instability in the use of point vortices for
computational fluid dynamics \cite{leonard}. Another problem is that
in three dimensions, the self-induced velocity of a curved point
vortex filament diverges logarithmically $\sim \ln(1/a)$ in the limit
of $a\to 0$ \cite{leonard}.  Since the optical intensity
decreases to zero near the center of an optical vortex, the quantum
pressure term, which determines the size of the vortex dark spot, can
no longer be ignored, and the optical vortex interactions should
differ markedly from point vortex interactions when their separation
is on the order of $a$.

To investigate the optical vortex dynamics when they are close to each
other, the fluid velocity is no longer an appropriate quantity to
study, because it diverges near a vortex center. The density, on the
other hand, approaches zero towards the center. This motivates us to
define an alternative finite quantity by multiplying the velocity and
the density,
\begin{align}
\mathbf{J} &\equiv \rho\mathbf{u}.
\end{align}
which is the fluid flux, or the momentum density. Simple calculations
show that the flux is indeed finite everywhere in an optical beam,
including the center of an optical vortex.  In terms of the flux, the
tensor dynamical equations now read \cite{grant}
\begin{align}
\parti{\rho}{\zeta} + \parti{J_i}{x_i} &= 0,
\label{flux_continuity}
\\
\parti{J_i}{\zeta} + \parti{}{x_j} \left(\frac{J_i J_j}{\rho} \right)
&=-\frac{1}{\mathcal{M}^2}\parti{}{x_i}\left(\frac{1}{2}\rho^2\right) -
\nonumber\\&\quad
\frac{1}{\mathcal{R}^2}\parti{}{x_j}\frac{1}{2}\left(
\parti{\sqrt{\rho}}{x_i}\parti{\sqrt{\rho}}{x_j}-
\sqrt{\rho}\parti{^2\sqrt{\rho}}{x_{i}\partial x_j}\right),
\label{flux_motion}
\end{align}
where 
$J_i$ is the $i$th component of $\mathbf{J}$, $\partial/\partial x_i$
is the $i$th spatial derivative, and repeated indices are implicitly
summed in the manner of Einstein's summation.  These equations have
the same form as the normalized Euler equations in the tensor form,
\begin{align}
\parti{\rho}{\zeta} + \parti{J_i}{x_i} &= 0,
\label{euler_flux_continuity}
\\
\parti{J_i}{\zeta} + \parti{}{x_j} \left(\frac{J_i J_j}{\rho} \right)
&=-\frac{1}{\mathcal{M}^2}\parti{P}{x_i},
\label{euler_flux_motion}
\end{align}
except the quantum pressure term in Eq.~(\ref{flux_motion}).  Hence,
in the flux representation, we have successfully avoided the problem
of divergent quantities.  Furthermore, Eq.~(\ref{flux_motion}), in
contrast to Eq.~(\ref{norm_bernoulli}), includes the correct
convective term.

The use of momentum density in the description of nonlinear optical
dynamics is more natural and appropriate than the use of velocity
in the Madelung transformation, as the dynamics ultimately evolves
according to the basic law of momentum conservation. As we shall
show next, when comparing the optical flux to the fluid flux,
the dynamics of optical vortex solitons are much more similar to
that of less singular fluid vortices than point vortices, and
the correspondence between nonlinear optics and Euler fluid
dynamics is still justified when $a$ is finite.

\subsection{Optical vortex solitons and vortex blobs}
In light of the fluid flux representation, one should therefore
compare the flux of an optical vortex soliton to the flux of a fluid
vortex. In an incompressible fluid, the density is constant, so the
flux is proportional to the velocity, and the flux at the center of a
point vortex has the same singular behavior as the velocity. Near a
vortex soliton, however, the flux is finite. Consider the example of a
single-charged vortex soliton. The flux near the center is given by
\begin{align}
\mathbf{J} &\propto \hat{\theta}r,\quad r << a,
\end{align}
which vanishes as $r\to 0$, as opposed to the divergence of
$\mathbf{J} \sim 1/r$ at the center of a point vortex.

\begin{figure}[htbp]
\centerline{\includegraphics[width=0.45\textwidth]{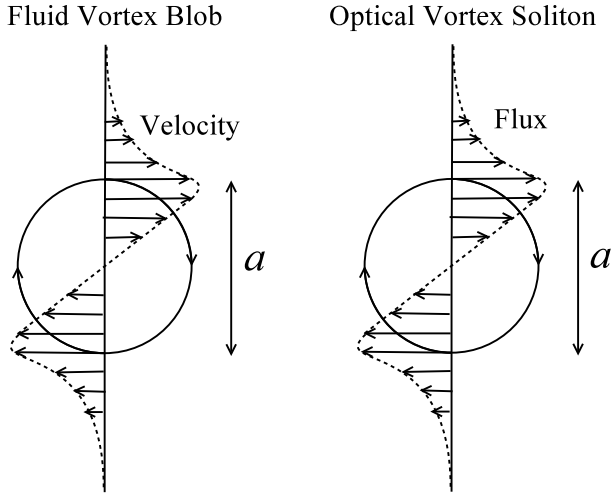}}
\caption{Sketches of velocity and flux of a vortex blob and
an optical vortex along a line across the center,
to illustrate the similarities
between the two in terms of the flux.}
\label{flux}
\end{figure}

Instead of comparing a vortex soliton to a point vortex, one should
hence compare the soliton to a \emph{vortex blob} \cite{leonard},
which has finite vorticity over a finite area. The vorticity
of a vortex blob filament is mathematically described by
\begin{align}
\pmb\omega(\mathbf{x},\zeta) &= 2\pi m_j \int d\mathbf{x}_j
\gamma(|\mathbf{x}-\mathbf{x}_j|)
\end{align}
where $\gamma$ is a vorticity distribution function for the filament.
The velocity near the center of a rectilinear vortex blob and
far away from the center is
\begin{align}
\mathbf{u} \propto \hat{\theta}r,\quad r << a,
\\
\mathbf{u} \propto \hat{\theta}\frac{1}{r}, \quad r >> a,
\end{align}
so in an incompressible fluid, the fluid flux of an optical vortex
soliton with size $a$ is the same as that of a vortex blob
with size $a$. See Fig.~\ref{flux} for a graphical illustration.
The dynamics of a vortex blob and that of a vortex soliton are also
extremely similar. For example, the rotation frequency $\Omega$ of two
like-charged vortex blobs approaches a constant $\propto 1/a^2$ when
their separation goes to zero. Numerical simulations of the nonlinear
Schr\"odinger equation also show that the rotation frequency of two
like-charged vortex solitons approaches a constant $\propto 1/a^2$ and
does not diverge like two point vortices \cite{aranson}. On the other
hand, the self-induced velocity of a curved vortex blob filament is
given by \cite{leonard}
\begin{align}
\parti{\mathbf{x}_i}{\zeta} &=
\frac{m_i\mathbf{b}}{2\rho_c}\ln\frac{\rho_c}{a}, 
\end{align}
where $\mathbf{b}$ is the unit binomial vector of the filament and
$\rho_c$ is the radius of curvative. The self-induced velocity of an
optical vortex soliton filament is proven to be exactly the same
\cite{pismen}. Hence, optical vortex solitons act as vortex blobs, and
a large number of solitons can simulate Euler fluid dynamics, much
like the popular discrete vortex blob method in computational fluid
dynamics \cite{leonard}.

\subsection{Numerical evidence of correspondence between
nonlinear optics and Euler fluid dynamics}
The most telling evidence of the correspondence between
nonlinear optics and Euler fluid dynamics is perhaps the numerical
fluid dynamics simulations using the nonlinear Schr\"odinger
equation by Nore \etal \cite{nore_physica,nore_shear}. Using
the nonlinear Schr\"odinger equation, Nore \etal numerical
demonstrated the Euler fluid dyanmics of a jet made of an array of
counter-rotating vortices, which exhibit sinuous and
varicose instabilities \cite{nore_physica}. In another
study, Nore \etal also demonstrated three-dimensional
shear flows and showed that numerically solving
nonlinear Schr\"odinger equation is a viable alternative
to Euler and Navier-Stokes equations for the numerical
study of shear flows \cite{nore_shear}. As nonlinear optical
propagation is governed by nonlinear Schr\"odinger equation,
The numerical experiments by Nore \etal show that nonlinear
optics should also be able to compute Euler fluid dynamics.

\section{Similarities between nonlinear Schr\"odinger dynamics and
Navier-Stokes fluid dynamics}
In the previous sections, we have shown the correspondence between
self-defocusing optical propagation and inviscid Euler fluid dynamics
via a variety of methods, including the Madelung transformation, the
incorporation of vorticity effect due to the ``phase slip''
phenomenon, the fluid flux representation, and the comparison between
optical vortex solitons and vortex blobs.  Even though viscosity plays
the role of a small parameter in most interesting fluid dynamics
problems, its effects are of paramount importance near a ``no-slip''
boundary and in the dissipation of eddies, in which cases the viscous
Navier-Stokes equations should be used.  In this section we shall
present evidence that the nonlinear Schr\"odinger equation exhibits
many of the same behaviors of viscous Navier-Stokes fluid dynamics,
and in each case, quantum pressure plays an analogous role to
viscosity.

The normalized Navier-Stokes equations in the flux representation are
given by
\begin{align}
\parti{\rho}{\zeta} + \parti{J_i}{x_i} &= 0,
\label{NS_continuity}
\\
\parti{J_i}{\zeta} + \parti{}{x_j} \left(\frac{J_i J_j}{\rho} \right)
&=-\frac{1}{\mathcal{M}^2}\parti{P}{x_i} +
\nonumber\\&\quad
\frac{1}{\mathcal{R}}
\parti{}{x_j}\left(\parti{u_i}{x_j}+\parti{u_j}{x_i}\right),
\label{NS_motion}
\end{align}
where the last term in Eq.~(\ref{NS_motion}) is the viscosity
term and $\mathcal{R}$ is called the Reynolds number, which describes the
relative strength of convection compared to viscosity,
\begin{align}
\mathcal{R} &= \frac{UL}{\nu},
\end{align}
where $U$ is the characteristic velocity of the fluid system, $L$ is
the characteristic length, and $\nu$ is the kinematic viscosity of the
fluid. Comparing the viscosity term in Eq.~(\ref{NS_motion}) with the
quantum pressure term in Eq.~(\ref{flux_motion}) via a dimensional
analysis would suggest that an analogous optical Reynolds number would
be defined as
\begin{align}
\mathcal{R} &= KW,
\label{opticalR}
\end{align}
where, to recall, $K$ is the characteristic optical phase gradient,
and $W$ is the characteristic size of the optical experiment
setup. The optical Reynolds number thus roughly measures the number of
optical vortices. In other words, if the optical Reynolds number
indeed corresponds to its fluid counterpart, then the quantization of
the optical vortices would play an analogous role to fluid viscosity.
This view seems to be echoed by other researchers in the field of
superfluids \cite{nore_PRL,nore_PF,stiessberger,vinen_review},
although we must stress that it is still an open problem as to what
extent the quantization effect resembles the viscous effect
\cite{vinen_review}.

\subsection{Zero-flux boundary conditions, boundary layers,
and boundary layer separation}
\begin{figure}[htbp]
\centerline{\includegraphics[width=0.45\textwidth]{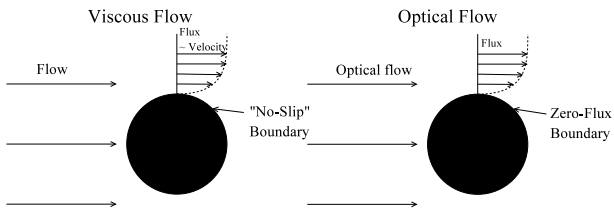}}
\caption{Comparison between a viscous boundary layer and
an optical boundary layer.}
\label{noslip}
\end{figure}

In classical fluid dynamics the ``no-slip'' boundary condition is most
commonly used, and restricts the total velocity and hence the total
flux to be zero at the boundary. For fluid flow above a surface, the
velocity shear introduced must be balanced by a viscous stress,
resulting in a boundary layer that connects the zero velocity at the
boundary to the flow velocity above the boundary in an asymptotic
expansion \cite{prandtl}. For the nonlinear Schr\"odinger equation,
the boundary condition of an impenetrable object can be specified by a
very low refractive index region, which restricts the optical
intensity to be zero at the surface \cite{berloff} due to total
internal reflection.  Even though the tangential velocity can have a
non-zero value at the surface, both the normal and tangential
components of the flux must be zero there. This can hence be viewed as
a zero-flux ``no-slip'' boundary condition.  An optical boundary layer
analogous to the viscous boundary layer in classical fluid dynamics is
also formed \cite{berloff}. See Fig.~\ref{noslip} for a graphical
illustration of the similarities between a viscous boundary layer and
an optical boundary layer.

For a viscous fluid flow past an obstacle, as the Reynolds number
increases, the boundary layer begins to separate and vorticity is
convected behind the obstacle. An analogy in the dynamics of the
nonlinear Schr\"odinger equation, in the form of vortex nucleation on
the boundary, is also predicted \cite{frisch}, and in the case of
large objects, the instability of the optical boundary layer also
depends on the optical Reynolds number $\mathcal{R}$ defined in
Eq.~(\ref{opticalR}) \cite{stiessberger}, much like the
viscous boundary layer separation.

\subsection{Dissipation of eddies}
Another important effect of viscosity is the dissipation of
small-scale structures in turbulence. An analogous effect in nonlinear
Schr\"odinger equation is the emission of sound waves two vortices are
close to each other \cite{aranson} and the generation of Kelvin waves
in the process of vortex line reconnections \cite{kozik}. The
radiation of acoustic energy in both cases must cause a damping of the
high-spatial-frequency convection within the optical beam, and the
effective Reynolds number is again estimated to be equal to the
optical Reynolds number \cite{nore_PRL,nore_PF,vinen_review}.

\subsection{K\'arm\'an vortex street}
The K\'arm\'an vortex street is a famous viscous fluid phenomenon, in
which alternate fluid vortices are emitted from the back of an
obstacle to the flow of a viscous fluid, when the Reynolds number
increases beyond a certain threshold \cite{thomson,williamson}. Using the
numerical vortex blob method, Chorin first simulated such a phenomenon
for a cylinder obstacle and obtained good agreement with experimental
data \cite{chorin}. Since an optical beam diffracting past a low
refractive index region would also emit optical vortices and the
vortices interact like vortex blobs in a self-defocusing medium, we
performed a numerical experiment of the nonlinear Schr\"odinger
equation to investigate if we would observe a similar phenomenon for
nonlinear optics.

\begin{figure}[htbp]
\centerline{\includegraphics[width=0.45\textwidth]{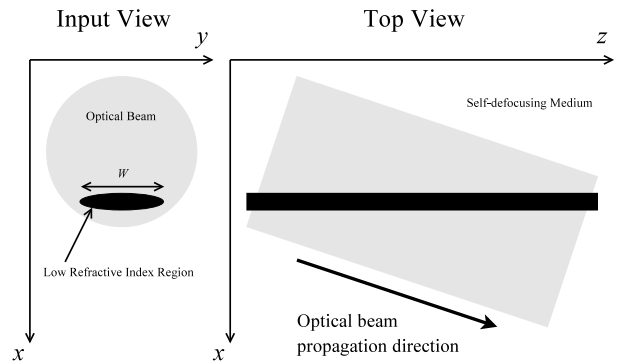}}
\caption{Setup of numerical experiment (not-to-scale).}
\label{setup}
\end{figure}

\begin{figure}[htbp]
\centerline{\includegraphics[width=0.45\textwidth]{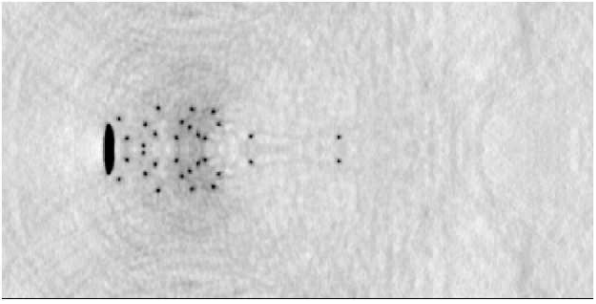}}
\caption{The intensity of the optical beam at a normalized
propagation distance $\zeta = 10$, for $\mathcal{M} = 0.4$ and
$\mathcal{R} = 12.8$. The dark ellipse is the low-refractive-index
region that acts as an impenetrable object. Optical vortex solitons
are seen to be created on the top and bottom side of the ellipse,
While the convection of the solitons behind the object resembles
the twin vortices behind an obstacle in a viscous fluid flow.}
\label{N4_mid_intensity}
\end{figure}

The numerical setup is sketched in Fig.~\ref{setup}. A big optical
beam is assumed to propagate at an angle to an ellipsoid cylinder,
with a refractive index much lower than the surroundings to act as an
impenetrable object, in a self-defocusing medium. The length of the
long axis of the ellipsoid cross section is assumed to be $W$, and the
short axis is assumed to be one-fifth of $W$ throughout the
simulations.  The two-dimensional nonlinear Schr\"odinger equation is
solved using the Fourier split-step method \cite{agrawal}, which
implies a periodic boundary condition for the optical beam. This
should not affect the qualitative behavior of the dynamics, if the
optical beam is much bigger than the object. In all of the
simulations, the Mach number $\mathcal{M}$ is fixed at 0.4, while the
optical Reynolds number $\mathcal{R}$ is varied. Figure
\ref{N4_mid_intensity} plots the intensity of the optical beam at a
normalized propagation distance $\zeta = 10$ for an optical Reynolds
number $\mathcal{R} = KW = 12.8$. Optical vortex solitons are created
on the top and bottom side of the low-refractive-index region, and
they interact in such a way that resembles the phenomenon of twin
vortices behind an obstacle in a low-Reynolds-number viscous fluid flow.

\begin{figure}[htbp]
\centerline{\includegraphics[width=0.45\textwidth]{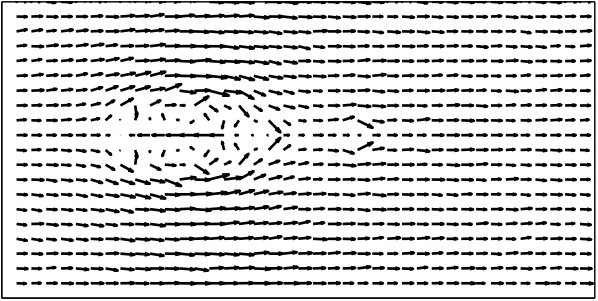}}
\caption{A vector plot of the flux $\mathbf{J}$ at
$\zeta = 10$, for $\mathcal{M} = 0.4$ and $\mathcal{R} = 12.8$,
which confirms the similarity between the numerically observed
dynamics and the phenomenon of twin vortices.}
\label{N4_mid_flux}
\end{figure}

\begin{figure}[htbp]
\centerline{\includegraphics[width=0.45\textwidth]{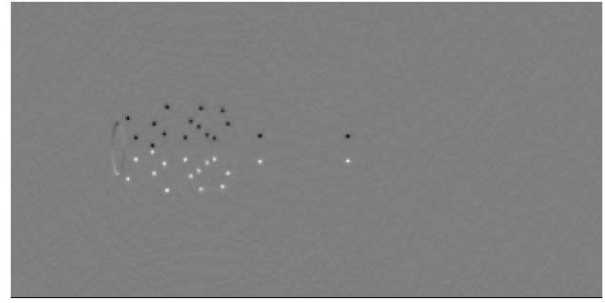}}
\caption{A plot of the momentum vorticity $\nabla\times\mathbf{J}$
at $\zeta = 10$, for $\mathcal{M} = 0.4$ and $\mathcal{R} = 12.8$.
A white dot indicates that the vortex has a positive topological charge and
a black dot indicates that the vortex has a negative charge. The
plot shows the similarity between the numerically observed dynamics
and the phenomenon of twin vortices.}
\label{N4_mid_vorticity}
\end{figure}

Figure \ref{N4_mid_flux} plots the flux $\mathbf{J} = 
(\psi^*\nabla\psi-\psi\nabla\psi^*)/2i$ and
Fig.~\ref{N4_mid_vorticity} plots the momentum vorticity
$\nabla\times\mathbf{J}$. Both plots confirm the similarity between
the numerically observed dynamics and the phenomenon of twin
vortices in a viscous fluid flow.

\begin{figure}[htbp]
\centerline{\includegraphics[width=0.45\textwidth]{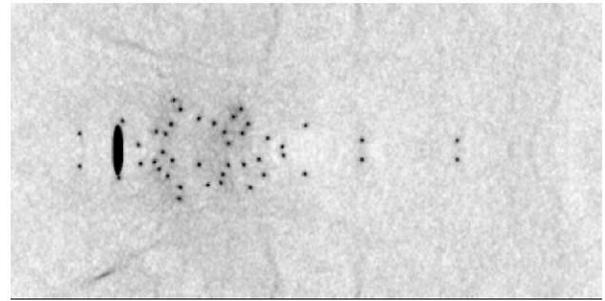}}
\caption{The intensity of the optical beam at a normalized
propagation distance $\zeta = 20$, for $\mathcal{M} = 0.4$ and
$\mathcal{R} = 12.8$. The qualitative dynamical behavior is
essentially unchanged from that shown in Fig.~\ref{N4_mid_intensity}.}
\label{N4_intensity}
\end{figure}

\begin{figure}[htbp]
\centerline{\includegraphics[width=0.45\textwidth]{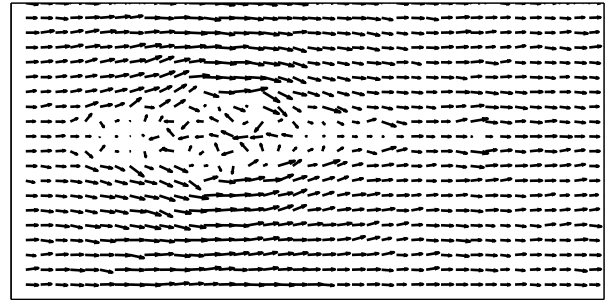}}
\caption{A vector plot of the flux $\mathbf{J}$ at
$\zeta = 20$, for $\mathcal{M} = 0.4$ and $\mathcal{R} = 12.8$.}
\label{N4_flux}
\end{figure}

\begin{figure}[htbp]
\centerline{\includegraphics[width=0.45\textwidth]{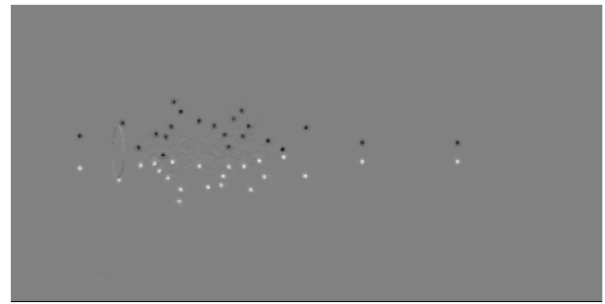}}
\caption{A plot of the momentum vorticity $\nabla\times\mathbf{J}$
at $\zeta = 20$, for $\mathcal{M} = 0.4$ and $\mathcal{R} = 12.8$.}
\label{N4_vorticity}
\end{figure}

Figures \ref{N4_intensity}, \ref{N4_flux}, and \ref{N4_vorticity}
plot the intensity, flux, and momentum vorticity of the optical beam
respectively, at a longer propagation distance $\zeta = 20$ for the
same parameters. The qualitative dynamical behavior of vortices
staying behind the object is essentially unchanged.

\begin{figure}[htbp]
\centerline{\includegraphics[width=0.45\textwidth]{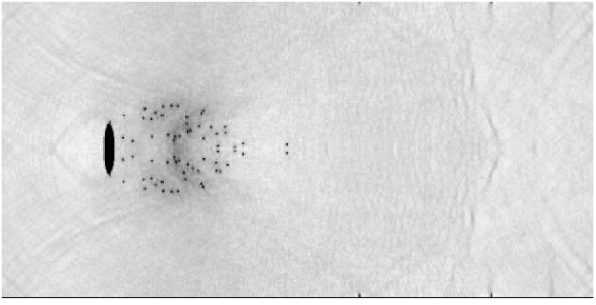}}
\caption{The optical intensity at $\zeta = 10$, for $\mathcal{M} = 0.4$
and $\mathcal{R} = 25.6$. The vortex solitons are observed to be
smaller, and the phenomenon of twin vortices is again observed.}
\label{N8_mid_intensity}
\end{figure}

\begin{figure}[htbp]
\centerline{\includegraphics[width=0.45\textwidth]{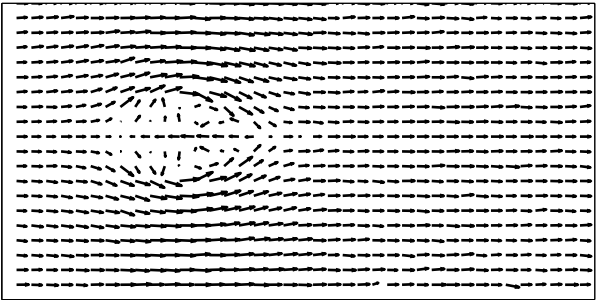}}
\caption{The flux $\mathbf{J}$ at $\zeta = 10$, for $\mathcal{M} = 0.4$
and $\mathcal{R} = 25.6$. }
\label{N8_mid_flux}
\end{figure}

\begin{figure}[htbp]
\centerline{\includegraphics[width=0.45\textwidth]{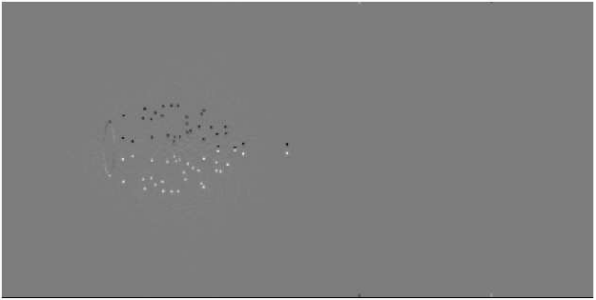}}
\caption{The momentum vorticity $\nabla\times\mathbf{J}$
at $\zeta = 10$, for $\mathcal{M} = 0.4$ and $\mathcal{R} = 25.6$.}
\label{N8_mid_vorticity}
\end{figure}

We now raise the Reynolds number to $\mathcal{R} = 25.6$ and perform
the numerical experiment again. As seen from
Figs.~\ref{N8_mid_intensity}, \ref{N8_mid_flux}, and
\ref{N8_mid_vorticity}, the optical vortex solitons become smaller and
more abundant, but at $\zeta = 10$ the phenomenon of twin vortices
behind an obstacle is again observed.

\begin{figure}[htbp]
\centerline{\includegraphics[width=0.45\textwidth]{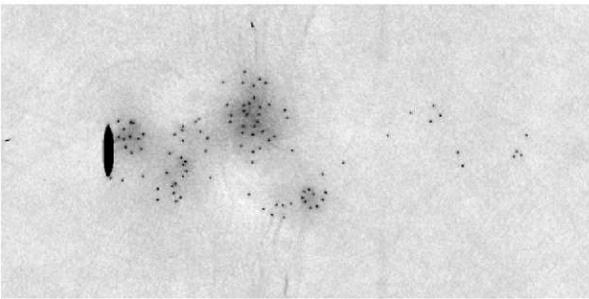}}
\caption{Optical intensity at $\zeta = 20$, for
$\mathcal{M} = 0.4$ and $\mathcal{R} = 25.6$. The twin
vortices become unstable and detach alternatively from
the object.}
\label{N8_intensity}
\end{figure}

\begin{figure}[htbp]
\centerline{\includegraphics[width=0.45\textwidth]{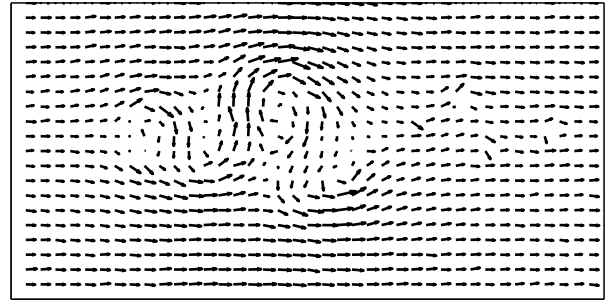}}
\caption{Flux at $\zeta = 20$, for $\mathcal{M} = 0.4$ and
$\mathcal{R} = 25.6$, which shows a flow pattern strongly
resembling the K\'arm\'an vortex street.}
\label{N8_flux}
\end{figure}

\begin{figure}[htbp]
\centerline{\includegraphics[width=0.45\textwidth]{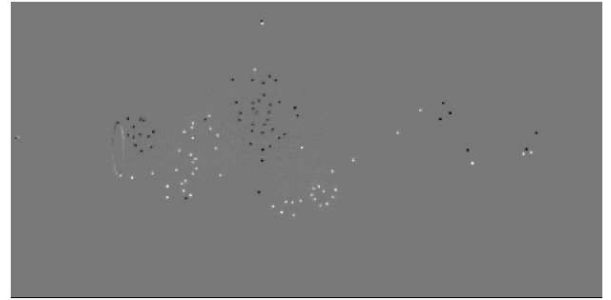}}
\caption{Vorticity at $\zeta = 20$, for $\mathcal{M} = 0.4$ and
$\mathcal{R} = 25.6$, which confirms that the alternate bunches
of vortices indeed have the right charges that resemble the
K\'arm\'an vortex street phenomenon.}
\label{N8_vorticity}
\end{figure}

At $\zeta = 20$, however, significant instability in the twin vortices
develops, such that the spatial symmetry between the upper plane and
the lower plane is broken, and alternative bunches of optical vortices
begin to be emitted from the back of the object. Figures
\ref{N8_intensity}, \ref{N8_flux}, and \ref{N8_vorticity} plot the
intensity, flux and vorticity at $\zeta = 20$ respectively, which
demonstrate a behavior strongly resembling the famous K\'arm\'an
vortex street phenomenon.

Due to computing power constraints, we are only able to simulate
low-Reynolds-number flows, which we do not expect to quantitatively
reproduce viscous fluid dynamics. We have to use an ellipsoid cylinder
in the numerical experiments, instead of the more conventional
circular cylinder, to artificially generate more optical vortices, and
the Mach number is a little too high for compressional waves not to
play a significant role in the dynamics. With all that said, using the
nonlinear Schr\"odinger equation, we are still able to qualitatively
demonstrate, for the first time to our knowledge, two well-known
viscous fluid phenomena, namely, the formation of twin vortices behind
an obstacle, and the symmetry-breaking instability of the twin
vortices that leads to the K\'arm\'an vortex street when the Reynolds
number is increased. Compared with previous claims of observing the
K\'arm\'an vortex street in nonlinear optics numerically
\cite{staliunas} or experimentally \cite{vaupel,molina}, our numerical
results demonstrate an unprecedented level of correspondence between
nonlinear optical dynamics and the K\'arm\'an vortex street
phenomenon, thanks to the presence of a much larger number of optical
vortices in our simulations.

\subsection{Kolmogorov turbulence}
The striking similarities between nonlinear optics and viscous fluid
dynamics are not limited to low-Reynolds-number two-dimensional
problems.  As the Reynolds number is further increased to the order of
a million, the viscous fluid flow enters a turbulent regime.  Since
this regime is highly chaotic, only statistical signatures can be
reproduced in a turbulent fluid flow. A well-known signature of
turbulence is the Kolmogorov energy spectrum \cite{kolmogorov},
derived under the assumption that a ``steady state'' is reached when
the macroscopic-scale fluid flow continuously generate finer spatial
structures via convection and viscosity dissipates the smallest
structures. As viscosity plays a significant role in the Kolmogorov
turbulence spectrum, it is surprising to see that numerical
simulations of the three-dimensional nonlinear Schr\"odinger equation
also reproduce the Kolmogorov spectrum at high Reynolds numbers, and
the vorticity dynamics of the ``superflow'' described by the nonlinear
Schr\"odinger equation resembles that of the viscous flow, in which
vortex reconnection events play a major role \cite{nore_PRL,nore_PF}.

The dissipation of the smallest spatial structures in a superflow is
speculated to be the Kelvin waves produced by the natural motion and
reconnections of vortex filaments \cite{vinen_review,kozik}, and the
corresponding Reynolds number is again speculated to be $\mathcal{R} =
KW$ \cite{vinen_review}. Numerical and theoretical analyses of the
so-called ``quantum turbulence'' exhibited by the nonlinear
Schr\"odinger equation all reveal striking similarities between
quantum and classical fluids, and it is argued that the study of
quantum turbulence could lead to a better understanding of turbulence
in normal fluids \cite{fisher}.

\section{The split-step method}
\begin{figure}[htbp]
\psfrag{Htotal}{$\hat{H}_{total} = \hat{H}_1 + \hat{H}_2 + ...$}
\psfrag{H1}{$\hat{H}_1$}
\psfrag{H2}{$\hat{H}_2$}
\psfrag{dz}{$\Delta\zeta$}
\centerline{\includegraphics[width=0.45\textwidth]{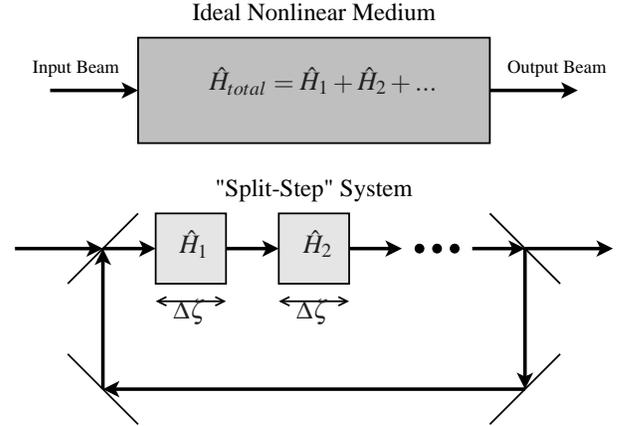}}
\caption{Sketch of a split-step optical system that approximates
the ideal nonlinear medium.}
\label{splitstep}
\end{figure}

While a nonlinear optical system shows promise for computing Euler and
Navier-Stokes fluid dynamics, it also poses serious technical
challenges. Ideally one would like to have a configurable nonlinear
material with low loss, anomalous group-velocity dispersion, high
defocusing nonlinearity, and three-dimensional co-propagating
boundaries. One may only be able to find separate materials or optical
devices, each of which performs only some of the functions. Moreover,
parasitic effects such as loss, two-photon absorption, and high-order
dispersion can be detrimental to the accuracy.  To combine different
devices and periodically compensate for parasitic effects, we hereby
propose the ``split-step'' method, the inspiration of which comes from the
numerical Fourier ``split-step'' method \cite{agrawal}. Consider the
general nonlinear Schr\"odinger equation,
\begin{align}
\parti{\psi}{\zeta} &= \sum_{n=1}^N \hat{H}_n \psi,
\end{align}
where propagation effects and boundary conditions are expressed in
terms of operators $\hat{H}_n$. The formal solution is
\begin{align}
\psi(\zeta+\Delta \zeta) &= \exp\Big(\int_\zeta^{\zeta+\Delta\zeta}
\sum_{n=1}^N \hat{H}_n d\zeta'\Big) \psi(\zeta).
\end{align}
But if $\Delta \zeta$ is much smaller than $1/H$ where $H$ is the
magnitude of the operators, by virtue of the Baker-Hausdorff formula
we have
\begin{align}
\psi(\zeta + \Delta\zeta) &=
\prod_{n=1}^N\exp\Big( \hat{H_n}\Delta\zeta
\Big)\psi(\zeta) +  O(H^2\Delta\zeta^2).
\end{align}
Each of the propagation effects can hence be applied separately to an
optical pulse, with a quadratic error term. A symmetrized version of
the split-step method can further reduce the error order,
\begin{align}
\psi(\zeta + \Delta\zeta) &=
\prod_{m=N}^1\exp\Big(\hat{H}_m\frac{\Delta\zeta}{2}\Big)
\prod_{n=1}^N\exp\Big( \hat{H_n}
\frac{\Delta\zeta}{2}\Big) \psi(\zeta) +
\nonumber\\&\quad  O(H^3\Delta\zeta^3).
\end{align}
The split-step method is not unlike the proof of a quantum computer
being able to simulate any quantum systems \cite{lloyd}. Whereas it is
difficult to find a quantum device that performs the exact Hamiltonian
of the quantum system of interest, it is possible to approximate the
Hamiltonian in small time slices. Similarly, in a metaphoric optical
computer, one can form a unit cell of a ``meta-material'' by combining
a slice of defocusing material, a slice of material with anomalous
group-velocity dispersion, a slice of ultrafast phase modulator to
apply the three-dimensional boundary conditions, and a gain medium to
compensate for loss.  The optical beam can loop through the unit cell
multiple times in a cavity, so that the outcome will approximate the
true solution as if we had an ideal medium. See Fig.~\ref{splitstep}
for a graphical illustration of the method.

The split-step method has the additional advantages that each
subsystem can be tunable and easily substituted with another material
or device, and the pulse evolution can be monitored more easily.  The
magnitude of each effect can be tuned by simply changing the
propagation length in each device, In exchange of configurability we
have sacrificed some accuracy due to discretization errors and
instability.  The computation speed may also be reduced by a large but
constant fraction, as the pulse may spend most of its time on simply
propagating from one device to the next and not performing the core
computation by nonlinear propagation. The split-step method,
however, does not detract from the inherent parallelism in the
computation, as the transverse dimensions are not discretized.

\section{Conclusion}
In conclusion, we have used a variety of theoretical
and numerical methods to show that self-defocusing optical
propagation has a converging correspondence with Euler fluid 
dynamics and a striking similarity with Navier-Stokes
fluid dynamics. We have numerically shown that the interactions
of a large number of optical vortex solitons are able to
simulate two well-known viscous fluid phenomena.
We have also proposed the split-step method,
a way of practically implementing the metaphoric optical
computer.

There are serious technical challenges if
a metaphoric computer is to become useful for computing fluid
dynamics, especially three-dimensional fluid dynamics problems,
as techniques for the complete specification and characterization
of the spatiotemporal optical field are still in their infancy.
The speed, configurability, and parallelism of a metaphoric optical
simulator nonetheless promise vast advantages over conventional
numerical simulations.

Since photons are quantum objects, optical propagation would also
inherently compute the quantum dynamics of bosons, and may therefore
be used as a metaphoric simulator of quantum fluids, such as
superfluids, superconductors, and Bose-Einstein condensates. In this
way the advantages of a metaphoric computer and those of a quantum
computer are combined, and only then the classical and quantum
computing capabilities offered by photons would truly be exhausted.
This extension of metaphoric optical computing will be a subject of
future work.

\section*{Acknowledgments}
We would like to acknowledge helpful and inspiring discussions
with Ravi Athale, Michael C. Cross, and Martin Centurion.
This work is supported by the Defense Advanced Research Projects
Agency (DARPA).

\end{document}